\documentclass[journal]{IEEEtran}
\IEEEoverridecommandlockouts
\usepackage{cite}
\usepackage{amsmath,amssymb,amsfonts}
\usepackage{algorithmic}
\usepackage{graphicx}
\usepackage{textcomp}
\usepackage{xcolor}

\def\BibTeX{{\rm B\kern-.05em{\sc i\kern-.025em b}\kern-.08em
    T\kern-.1667em\lower.7ex\hbox{E}\kern-.125emX}}
\begin{document}

\title{Modeling and Analysis of Switched-Capacitor Converters as a Multi-port Network for Covert Communication\\
}

\author{Yerzhan~Mustafa,~\IEEEmembership{Graduate~Student Member,~IEEE,}
        and~Selçuk~Köse,~\IEEEmembership{Member,~IEEE}% <-this % stops a space
\thanks{Y. Mustafa and S. Köse are with the Department
of Electrical and Computer Engineering, University of Rochester, Rochester,
NY, 14627, USA. E-mails: (yerzhan.mustafa@rochester.edu, selcuk.kose@rochester.edu).}% <-this % stops a space
\thanks{This work is supported in part by the NSF Award under Grant CNS-1929774.}
}
%\author{\IEEEauthorblockN{Yerzhan Mustafa}
%\IEEEauthorblockA{\textit{Department of Electrical and Computer Engineering} \\
%\textit{University of Rochester}\\
%Rochester, NY \\
%yerzhan.mustafa@rochester.edu}
%\and
%\IEEEauthorblockN{Selçuk Köse}
%\IEEEauthorblockA{\textit{Department of Electrical and Computer Engineering} \\
%\textit{University of Rochester}\\
%Rochester, NY \\
%selcuk.kose@rochester.edu}
%}

\maketitle

\begin{abstract}
Switched-capacitor (SC) DC-DC voltage converters are widely used in power delivery and management of modern integrated circuits. 
Connected to a common supply voltage, SC converters exhibit cross-regulation/coupling effects among loads connected to different SC converter stages due to the shared components such as switches, capacitors, and parasitic elements. 
The coupling effects between SC converter stages can potentially be used in covert communication, where two or more entities (\textit{e.g.}, loads) illegitimately establish a communication channel to exchange malicious  information stealthily. 
To qualitatively analyze the coupling effects, a novel modeling technique is proposed based on the multi-port network theory. 
The fast and slow switching limit (FSL and SSL) equivalent resistance concepts are used to analytically determine the impact of each design parameter such as switch resistance, flying capacitance, switching frequency, and parasitic resistance. 
A three-stage 2:1 SC converter supplying three different loads is considered as a case study to verify the proposed modeling technique.
\end{abstract}


\section{Introduction}
\IEEEPARstart{C}{ommunication} channels can be classified into two types such as covert and overt. 
An overt channel is legitimate and can be obvious to other entities, whereas a covert channel is established between malicious entities to potentially enable transmission of unauthorized content.
An example of such content could be sensitive information such as a secret key that is used for encryption and decryption algorithms. 
The entities could correspond to hardware components  (\textit{e.g.}, cores of the processor) and/or software applications.
In a communication channel, the information is transmitted from one entity to another. 
The transmitting and receiving ends are often referred to as source and sink, respectively. 

The communication media can be a hardware or software resource that is shared between the source and sink. 
For instance, the covert channel can be established via the memory cache that is shared between threads in a simultaneous multithreading processor \cite{percival2005cache,wang2006covert}, thermal sensors that measure the temperature of a processor's core \cite{masti2015thermal}, clock skew that can be affected by temperature variations \cite{murdoch2006hot}, memory bus \cite{wu2014whispers}, random number generator \cite{evtyushkin2016covert}, magnetic field sensors \cite{matyunin2016covert} and USB charging cable \cite{spolaor2017no} on mobile devices, and general purpose GPUs \cite{naghibijouybari2017constructing}.

A covert communication can also be established via the power management units, which control voltage level and operating frequency. 
A covert channel attack among FPGAs, CPUs, and GPUs that are connected to the same power supply unit is  proposed in~\cite{giechaskiel2020c}. 
By modulating the core frequency with dynamic frequency scaling (DFS) technique, a covert channel has been created in a multi-core platform \cite{alagappan2017dfs}. 
The use of dynamic voltage and frequency scaling (DVFS) allowed to create a covert channel in a TrustZone-enabled system-on-chip (SoC) \cite{bossuet2018dvfs}. 
Information leakage is enabled by infecting power management unit with a hardware Trojan in the multi-core SoC in~\cite{islam2018pmu}. 
The power budget constraint \cite{khatamifard2019powert} and switching noise modulation \cite{wang2019exploring} of power delivery network (PDN) can also create the covert communication between cores of the processor.

%The SC converter is a voltage converter that changes one DC voltage level to another one by periodically switching charging and discharging phases of capacitor(s). These converters are widely used in power management IC. 

Similar to the aforementioned circuit and system components that are shared by multiple entities and used for covert communication, on-chip voltage regulators and PDN are also shared by multiple entities and can therefore be used as a communication medium.
SC converters are widely used in as on-chip voltage regulators in modern integrated circuits (ICs).
One of the primary parameters of an SC converter is the equivalent resistance, which characterizes the average ohmic losses. 
The equivalent resistance is a function of the switch resistance, flying capacitance, switching frequency, duty cycle, and parasitic elements such as equivalent series resistance (ESR) and equivalent series inductance (ESL).
Several techniques have been proposed for analytical modeling of the equivalent resistance. 
One of the most common techniques are slow switching limit (SSL) and fast switching limit (FSL) approximations \cite{seeman2009design}. 

In SSL approximation, the switching period is quite large (\textit{i.e.}, the switching frequency is small) such that the capacitors in an SC converter have enough time to completely charge or discharge within one switching sub-interval. 
In this case, the switch resistance can be approximated to be zero. 
In FSL approximation, the switching period is quite small (\textit{i.e.}, the switching frequency is large), allowing the capacitors to behave as a constant DC voltage sources because there is not enough time to charge or discharge within one switching sub-interval. 
There exists more complex/accurate calculation technique of equivalent resistance \cite{mustafa2018scc}. 
However, the additional computational complexity of the equivalent resistance derivation makes these techniques typically infeasible.
%the trade-off is in the computational complexity (\textit{i.e.}, as the circuit becomes larger, it is harder to calculate the expression of equivalent resistance).

SC converters may share certain components such as switches and capacitors among different entities. 
Coupling and cross-regulation effects of one entity can therefore be observed from another entity. 
The variations in one of the outputs of an SC converter can cause an observable change in the performance of the another output and vice versa. 
The coupling effects are investigated in dual- and multi-output SC converters by extending the notion of equivalent resistance in~\cite{mustafa2018dual,delos2014modeling}. 

This paper proposes a modeling technique of SC converters that can be used to characterize the covert communication capacity among different circuit entities that share a common PDN. 
By using the concept of multi-port network, the cross-regulation/coupling effects among different load circuitry, which are powered by different stages of SC converters, are characterized. 
The following are key contributions of this work.

\begin{itemize}
    \item Introduction and analysis of a novel covert communication channel, which can be established between SC converter stages sharing the same PDN. 
    \item Discussion of practical threat model that could exploit this vulnerability on real devices.
    \item Analytical- and simulation-based modeling techniques that account for the coupling/cross-regulation effects between SC converter stages and quantitatively characterize the bandwidth of the covert communication channel.
    
    %\selcuk{strength or bandwidth?} \yerzhan{\{Here I meant strength. Coupling characterize output voltage variation, which is a difference in voltage amplitude of between bit 0 and bit 1. Bandwidth depends on the response time of SC converter assuming that we want to preserve the highest possible coupling.\}} 
    \item Demonstration of these modeling techniques with a case study and analysis of the SC converter design parameters (\textit{e.g.}, location on chip, switching frequency, off-chip resistance) that could either increase or decrease the bandwidth of covert communication.
    \item Verification of the proposed analytical modeling technique with comprehensive simulation results. 
\end{itemize}

This paper is organized as follows. 
The threat model which can be used to establish the covert communication between cores of the processor is presented in Section \ref{threat_model}. A generalized multi-port network modeling technique that can be used to analyze SC converter covert communication with both analytical calculations and simulations is proposed in Section \ref{generalized_multi-port_network}. 
An example case study to explain the proposed modeling technique on a three-stage 2:1 SC converter is offered in Section \ref{case_study}. 
The accuracy of analytical modeling technique is verified with extensive simulations in Section \ref{simulation_covert_communication}. 
The conclusions are drawn in Section \ref{conclusion}.

\section{Threat Model}\label{threat_model}

\begin{figure}[htbp]
	\centering
	\includegraphics[width=0.4\textwidth]{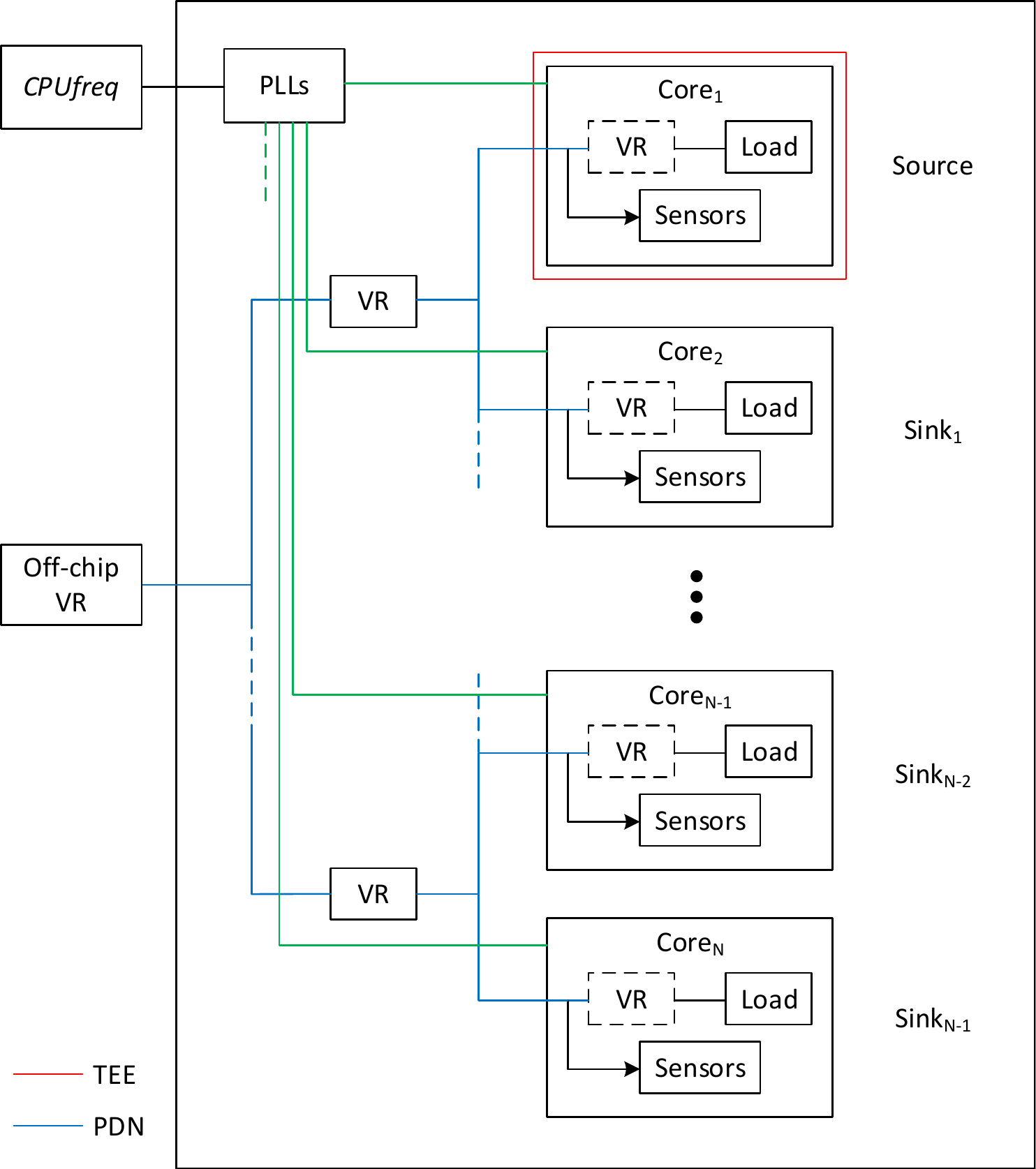}
	\caption{Threat model for covert communication among different cores.}
	\label{fig:threat_model}
\end{figure}

In the threat model (Fig. \ref{fig:threat_model}), two or more entities (\emph{e.g.}, cores, applications, or programs) such as source and sink would like to establish the communication channel. 
Assume that the source entity has access to sensitive information while the sink entity has access to the network. 
A common practice to protect the sensitive information is to locate the source in a Trusted Execution Environment (TEE). 
No overt communication channel can therefore be established between the source and sink.

Due to the shared PDN, it is possible to establish the covert communication channel between cores \cite{khatamifard2019powert, wang2019exploring}. 
As can be observed in Fig. \ref{fig:threat_model}, the power can be delivered and regulated by on-chip voltage regulators (VRs). 
Depending on the architecture of the microprocessor, the VRs can be connected to the cores in different configurations.
%there are different configurations in which VRs can be connected to cores. 
Certain designs can have a dedicated VR for each core (highlighted in dashed line in Fig. \ref{fig:threat_model}). 
For instance, Qualcomm Snapdragon 800 supports per core DVFS, \textit{i.e.}, there is one VR per each core to control voltage \cite{qualcomm_2021}. 
Alternatively, some implementations can have less number of VRs such that cores are organized in clusters, where each cluster have a dedicated VR. 
As an example, the ARM big.LITTLE architecture has two clusters of high performance and low power cores~\cite{ltd.}. 

A communication signal can be sent by the source through the PDN by varying the load from heavy to light conditions. 
This can be implemented by pinning the application to a specific core that could run in either high performance or idle (sleep) modes to encode logical 1 and 0, respectively. 
Because of the coupling effects (\textit{i.e.}, shared components in PDN), the variations in the load at the source side propagates to other cores. 
Therefore, the sensitive information can be leaked from the source and sensed by the sink(s).

The variations in load primarily affect the core voltage. 
In order to decode the signal sent from source, the sink should be able to measure the voltage level with certain sensors. 
One possible way to measure the variations in voltage could be performed by using the critical path monitor (CPM) sensors, which can be found in, \textit{e.g.}, IBM POWER7+ multicore processors \cite{drake2007distributed,drake2013single}.
The CPM sensors detect changes in the timing margin of guardband scheduling. 
The output of CPM sensors is typically in near-linear relationship with core voltage \cite{zu2015adaptive}. 
In POWER7+ processor, each core has five CPM sensors, which can be accessed with IBM AMESTER software \cite{floyd2011introducing}.

The operating frequency of each core can have a significant impact on the operation characteristics of respective VRs such as the output voltage and switching frequency.
As a result, it is important to select the core frequency in such a way to maximize the bandwidth of the communication channel. 
The operating frequency of each core is usually controlled by phase-locked loops (PLLs), as shown in Fig. \ref{fig:threat_model} and the frequency of each core can be controlled by software. 
\textit{E.g.}, Operating Systems (OS) based on Linux have \textit{CPUfreq} software, which can set the frequency for each core. 
This software can be accessed with the kernel privileges to implement the attack \cite{noubir2020towards,qiu2019voltjockey,islam2018pmu, alagappan2017dfs}. 
Additionally, the location of the source and sink significantly affects the bandwidth of communication signal due the different level of coupling between cores. 
A novel modeling technique that accounts for the location of source and sink, core frequency, and other parameters so as to increase the bandwidth of communication signal is proposed in this paper.

\section{Generalized Multi-Port Network Model}\label{generalized_multi-port_network}
\begin{figure}[t]
	\centering
	\includegraphics[width=0.35\textwidth]{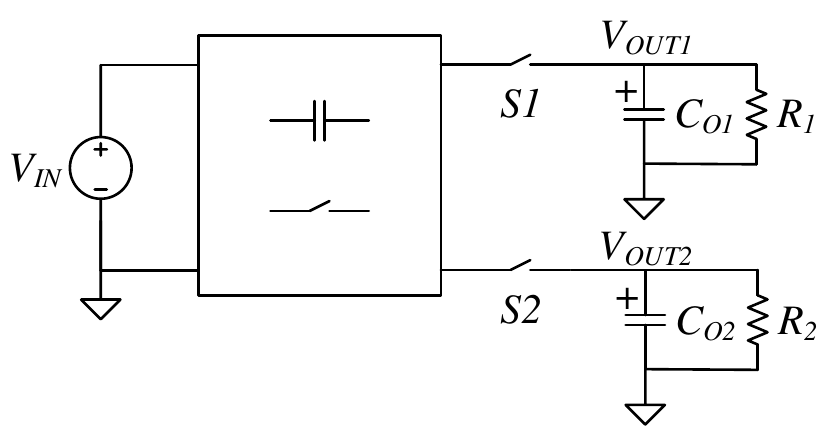}
	\caption{General structure of a dual-output SC converter.}
	\label{fig:general_structure_dual}
\end{figure}

A dual-output SC converter is a type of DC-DC converter that provides two outputs by using shared components such as flying capacitors and switches, as shown in Fig. \ref{fig:general_structure_dual}. 
The component sharing is utilized to minimize the area and power overhead in dual-output SC converters especially in the discrete circuits. 
The main disadvantage of this technique is, however, the cross-regulation/coupling effects between output terminals where the load behavior of one output affects the voltage at the other output. 
In this work, the cross-regulation/coupling effects are analyzed and their impact on the formation of a possible covert communication channel among SC converter stages is characterized.

In IC design, the dual-output SC converters are rarely used because the component sharing is not needed due to scaling. 
Instead, multitude of single-output SC converters - so called multi-phase SC converters - are used in parallel to supply loads. 
For example, \cite{lu201520} proposed a 123-phase SC DC-DC converter-ring for microprocessor applications.

\begin{figure}[t]
	\centering
	\includegraphics[width=0.4\textwidth]{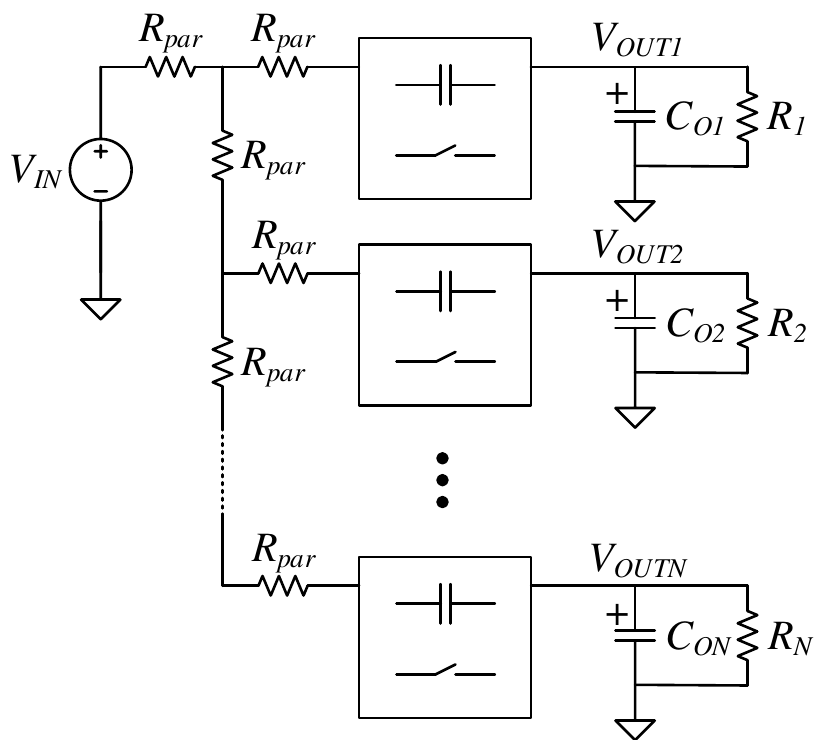}
	\caption{Multi-phase SC converter with $N$ number of stages supplying $N$ different loads.}
	\label{fig:N-port_network}
\end{figure}

A dual-output SC converter model that accounts for those cross-regulation/coupling effects is proposed in~\cite{mustafa2018dual}, which uses the concept of two-port network (system).
Similar to a dual-output SC converter, a multi-phase SC converter with two stages can be viewed as a two-port network, where each stage supplies a separate load. 
Moreover, this modeling can be generalized to the multi-port network, where the multi-phase SC converter with $N$ number of stages is providing power to $N$ different loads, as illustrated in Fig. \ref{fig:N-port_network}. 
The parasitic resistance, $R_{par}$, is added to account for the location of SC converter stage and its load with respect to the input voltage source. 
The equal parasitic resistance is chosen to simplify further derivations. 
Please note that $R_{par}$ can have different values.

\subsection{Generalized formulas}
The multi-phase SC converter with $N$ number of stages that provides power to $N$ number of loads can be modelled as an $N$-port network, where the average output voltages and load currents can be written as
\begin{equation}
\small
\begin{bmatrix}
V_{OUT1}\\
V_{OUT2}\\
\vdots\\
V_{OUTN}
\end{bmatrix}
=
\begin{bmatrix}
V_{TR1}\\
V_{TR2}\\
\vdots\\
V_{TRN}
\end{bmatrix}
-
\begin{bmatrix}
R_{11} & R_{12} & \cdots & R_{1N}\\
R_{21} & R_{22} & \cdots & R_{2N}\\
\vdots & \vdots & \ddots & \vdots\\
R_{N1} & R_{N2} & \cdots & R_{NN}
\end{bmatrix}
\begin{bmatrix}
I_{OUT1}\\
I_{OUT2}\\
\vdots\\
I_{OUTN}
\end{bmatrix}
,
\label{eq_generalized_multi_port}
\end{equation}

\noindent
where $V_{OUTi}$ is the average output voltage and $V_{TRi}$ is the target (no load) output voltage of $i^{th}$ terminal. 
For the load resistance of the $i^{th}$ terminal, $R_i$, the average output currents are determined by $I_{OUTi}=V_{OUTi}/R_i$.

In R-parameters matrix, intuitively, the $R_{ij}$ represents the amount of coupling from the $j^{th}$ load to the $i^{th}$ load. 
Therefore, the non-diagonal elements are of the primary parameters to analyze the coupling effects. 
%This can be used quantify the covert communication between two loads.
%
It should be noted that the dimensions of matrices in (\ref{eq_generalized_multi_port}) depend only on the number of loads. 
This means that the number of SC converter stages does not have to be $N$ (\textit{i.e.}, several converters may supply the same load). 
Additionally, the location as well as the number of the input voltage sources can be different, which also does not change the dimension of the matrices. 
The change in the input voltage location and/or the number of SC converter stages and input voltage sources will only change the values of the elements of R-parameters matrix.

\subsection{Analytical modeling}\label{analytical_modeling_general}
The R-parameters can be determined without FSL and SSL equivalent resistance approximations~\cite{mustafa2018dual}. 
Although these formulations in~\cite{mustafa2018dual} are accurate across the switching frequency range, the trade-off is in the computational complexity. 
As the circuit becomes larger, the calculation of R-parameters will be more difficult due the high-order RC circuits \cite{mustafa2018scc}. 
In this work, the analytical expressions of R-parameters are determined using FSL and SSL equivalent resistances \cite{seeman2009design}.

\begin{figure}[t]
	\centering
	\includegraphics[width=0.4\textwidth]{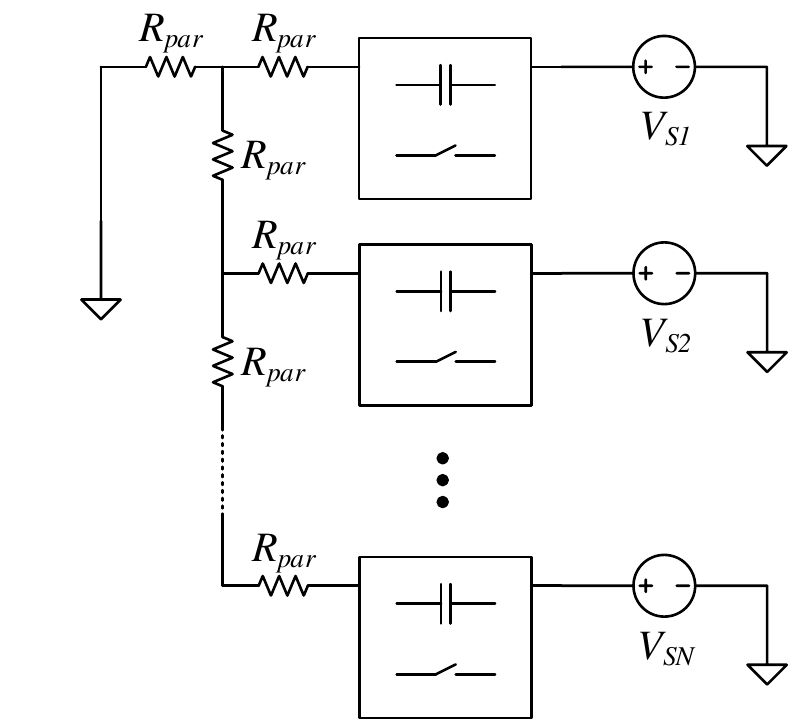}
	\caption{Analytical modeling of R-parameters for an $N$-port network.}
	\label{fig:r_par_calculation_N-port}
\end{figure}

To determine the analytical expressions of R-parameters, one should short-circuit the input voltage source and substitute loads with controllable voltage sources ($V_{Si}$). 
The complete analysis setup is shown in Fig. \ref{fig:r_par_calculation_N-port}.
In this setup, the effect of the filter/output capacitors is neglected, which is an appropriate assumption because the filter capacitors are usually much larger that the flying capacitors. 
As a result, (\ref{eq_generalized_multi_port}) can be re-written as

\begin{equation}
\begin{bmatrix}
I_{OUT1}\\
I_{OUT2}\\
\vdots\\
I_{OUTN}
\end{bmatrix}
=
-
\begin{bmatrix}
Y_{11} & Y_{12} & \cdots & Y_{1N}\\
Y_{21} & Y_{22} & \cdots & Y_{2N}\\
\vdots & \vdots & \ddots & \vdots\\
Y_{N1} & Y_{N2} & \cdots & Y_{NN}
\end{bmatrix}
\begin{bmatrix}
V_{S1}\\
V_{S2}\\
\vdots\\
V_{SN}
\end{bmatrix}
.
\label{eq_multi_port_calculation}
\end{equation}
Each element of the Y-parameters matrix can be found as
\begin{equation}
Y_{ij}=- \frac{I_{OUTi}}{V_{Sj}}; V_{Sk}=0,  k\neq j \quad (i,j,k \in \{1,2,...,N\}),
\label{eq_Yij}
\end{equation}

\noindent
where $V_{Sj}$ is the voltage source of arbitrary value $V$ connected to the $j^{th}$ output terminal and $V_{Sk}$ is a voltage source of zero value connected to the remaining output terminals.
In this way, the Y-parameters are determined, which then can be transformed into R-parameters by inversion as
\begin{equation}
\begin{bmatrix}
R_{11} & R_{12}\\
R_{21} & R_{22}
\end{bmatrix}
=
\begin{bmatrix}
Y_{11} & Y_{12}\\
Y_{21} & Y_{22}
\end{bmatrix}^{-1}.
\label{eq_RandY_parameters}
\end{equation}

Since the value of voltage sources are controlled and known, the main goal is to find the output current expressions. 
This can be achieved by assuming either SSL or FSL operation modes based on the switching frequency. In Section \ref{analytical_modeling}, examples of FSL and SSL Y-parameters calculation are provided.

\subsection{Simulation-based modeling}\label{sim_modeling}
In Section \ref{analytical_modeling_general}, the input voltage source was short-circuited to determine R-parameters matrix. 
This is done to eliminate $V_{TRi}$ parameters. 
For such scenario, switches are assumed to be ideal because during the R-parameters measurement, SC converter stages are assumed to operate in the same way as in normal condition (\textit{i.e.}, when the input voltage is connected and the  switches operate in the correct order). 
%\selcuk{what do you mean by normal condition here?} \yerzhan{\{Normal means switches in the converter operate correctly. If we short-circuit the input voltage source, then Vgs is different. From that, the non-ideal switch could not switch correctly and SC converter does not operate correctly. Therefore, we assume switches are ideal and do not depend on the Vgs.\}}
If one uses non-ideal MOSFETs, the gate-to-source voltage difference would have a limited amount of impact on the operation of SC converter stages, potentially increasing the error in the analytical derivations. 

\begin{figure}[t]
	\centering
	\includegraphics[width=0.4\textwidth]{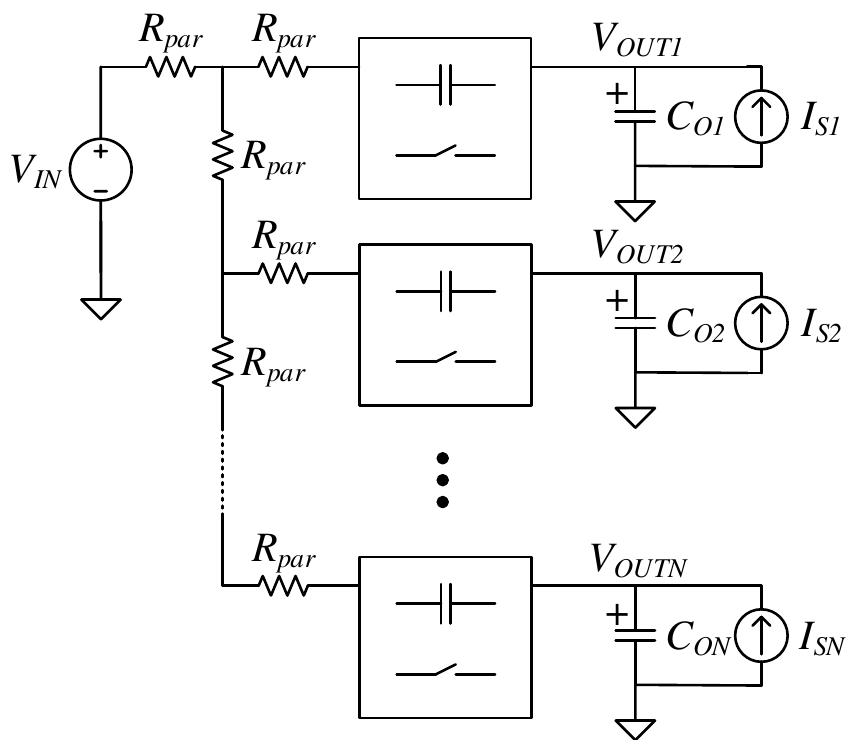}
	\caption{R-parameter measurements in simulation for $N$-port network.}
	\label{fig:r_par_measurement_N-port}
\end{figure}

It is possible to measure R-parameters without short-circuiting the input voltage source. 
In this way, non-ideal switches can be used during the measurement. 
This will require one more additional measurement.
To determine the R-parameters matrix, the current sources are added in place of load resistors. 
The setup is shown in Fig. \ref{fig:r_par_measurement_N-port}. 
Note that the input voltage source is not-short-circuited. 
Moreover, the effect of filter capacitors is accounted in this setup.

With this configuration, the average output voltages can be expressed as
\begin{equation}
\small
\begin{bmatrix}
V_{OUT1}\\
V_{OUT2}\\
\vdots\\
V_{OUTN}
\end{bmatrix}
=
\begin{bmatrix}
V_{TR1}\\
V_{TR2}\\
\vdots\\
V_{TRN}
\end{bmatrix}
+
\begin{bmatrix}
R_{11} & R_{12} & \cdots & R_{1N}\\
R_{21} & R_{22} & \cdots & R_{2N}\\
\vdots & \vdots & \ddots & \vdots\\
R_{N1} & R_{N2} & \cdots & R_{NN}
\end{bmatrix}
\begin{bmatrix}
I_{S1}\\
I_{S2}\\
\vdots\\
I_{SN}
\end{bmatrix}
,
\label{eq_generalized_multi_port_simulation}
\end{equation}
By setting the value of all of the current sources to zero, the target voltages $V_{TRi}$ can be determined as
\begin{equation}
\begin{bmatrix}
V_{TR1}\\
V_{TR2}\\
\vdots\\
V_{TRN}
\end{bmatrix}
=
\begin{bmatrix}
V_{OUT1}\\
V_{OUT2}\\
\vdots\\
V_{OUTN}
\end{bmatrix}
.
\label{eq_target_VOUT}
\end{equation}
Each element of the R-parameters matrix can then be found as
\begin{equation}
R_{ij}=\frac{V_{OUTi}-V_{TRi}}{I_{Sj}}; I_{Sk}=0,  k\neq j \; (i,j,k \in \{1,2,...,N\}),
\label{eq_Rij}
\end{equation}

\noindent
where $I_{Sj}$ is the current source of arbitrary value $I$ connected to the $j^{th}$ output terminal and $I_{Sk}$ is a current source of zero value connected to the remaining output terminals.
The elements of R-parameters matrix can be determined by changing the values of the current sources $N$ times and measuring the output voltage values.

In case the current sources are not feasible to be realized, load resistors can be used instead of the current sources. 
By controlling the load resistance value, the load current can be controlled. 
For example, to model $I_{Si}=0$ the load resistance needs be set to an extremely large value. 
Similarly, to model an arbitrary current source, the load resistance should be set to a fixed arbitrary value. 
Accordingly, the load current can be written as $I_{Si}=-I_{OUTi}$.

\section{Case Study: Modeling of Three-Stage 2:1 SC Converter}\label{case_study}

\begin{figure}[t]
	\centering
	\includegraphics[width=0.4\textwidth]{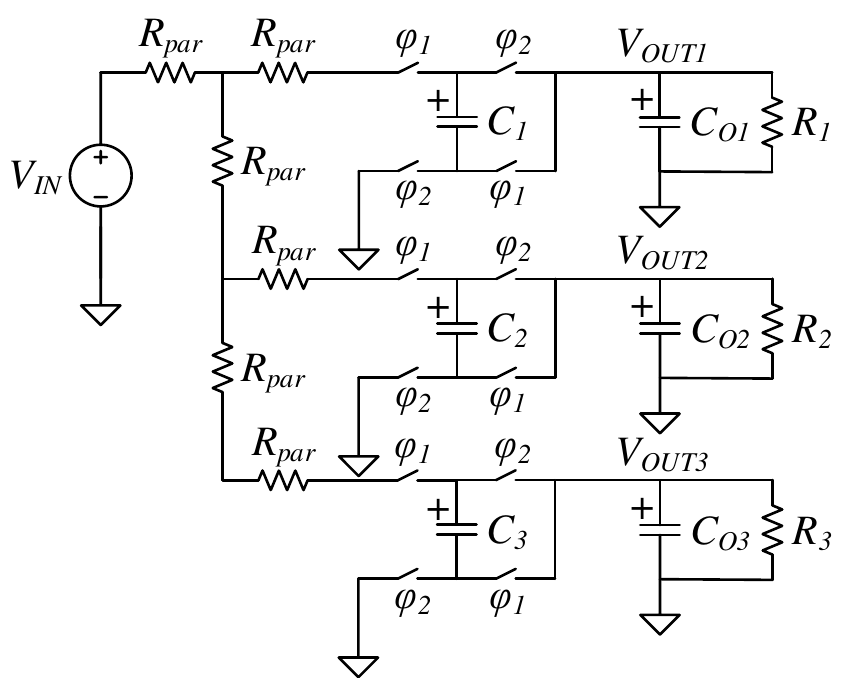}
	\caption{Three-stage 2:1 SC converter supplying three different loads.}
	\label{fig:3-port_network}
\end{figure}

In this case study, a three-stage 2:1 SC converter is considered, as shown in Fig. \ref{fig:3-port_network}. 
Each stage of this circuit operates in two phases (charge and discharge) while producing $N=3$ output voltages that are equal to half of the input voltage. 

\subsection{Analytical modeling}\label{analytical_modeling}
\subsubsection{FSL}

\begin{figure}[t]
	\centering
	\includegraphics[width=0.45\textwidth]{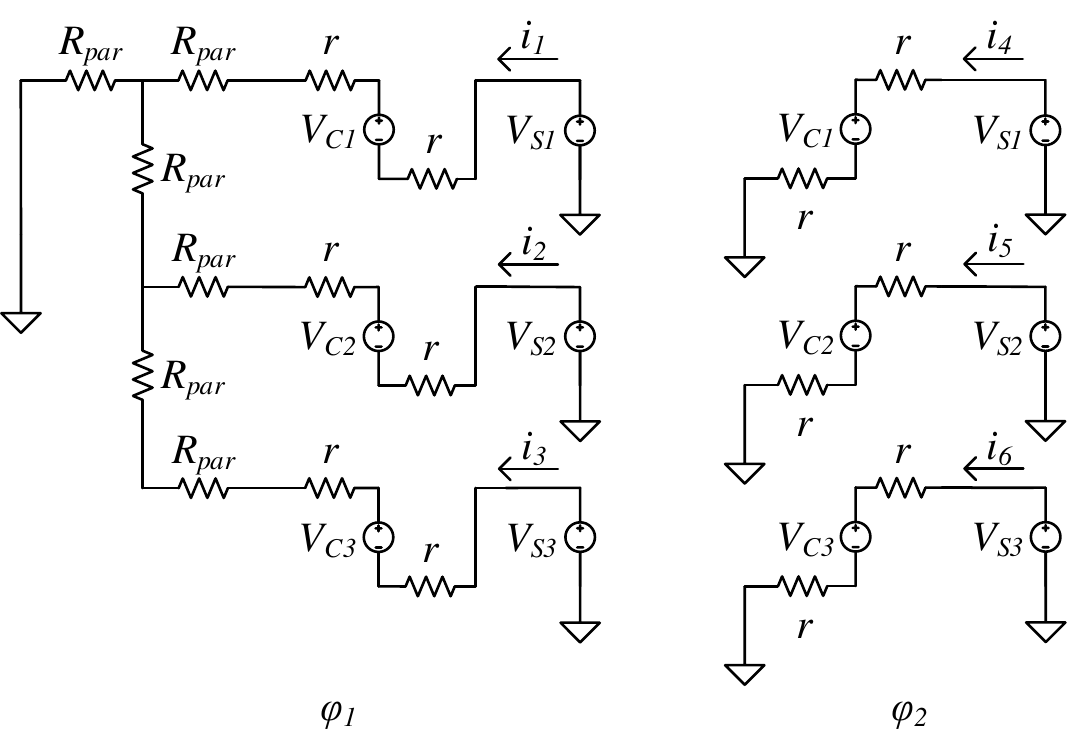}
	\caption{FSL Y-parameters calculation for a three-stage 2:1 SC converter supplying three different loads.}
	\label{fig:FSL_3-port_network}
\end{figure}

The circuit diagram used to determine FSL Y-parameters is shown in Fig. \ref{fig:FSL_3-port_network}. 

In Fig. \ref{fig:FSL_3-port_network}, there are nine unknown parameters such as $V_{C1-3})$ and $i_{1-6}$. 
To determine their expressions, a set of nine linear equations need to be constructed and solved. 
This can be achieved by applying Kirchhoff's Voltage Law (KVL) and charge balance equations, which are given by
\begin{equation}
2i_1r-V_{C1}+i_1R_{par}+(i_1+i_2+i_3)R_{par}=V_{S1};
\label{eq_KVL1}
\end{equation}

\begin{equation}
2i_2r-V_{C2}+i_2R_{par}+(i_2+i_3)R_{par}+(i_1+i_2+i_3)R_{par}=V_{S2};
\label{eq_KVL2}
\end{equation}

\begin{equation}
2i_3r-V_{C3}+2i_3R_{par}+(i_2+i_3)R_{par}+(i_1+i_2+i_3)R_{par}=V_{S3};
\label{eq_KVL3}
\end{equation}

\begin{equation}
\int_{0}^{T/2} i_1 \,dt = \int_{0}^{T/2} i_4 \,dt;
\label{eq_charge1}
\end{equation}

\begin{equation}
\int_{0}^{T/2} i_2 \,dt = \int_{0}^{T/2} i_5 \,dt;
\label{eq_charge2}
\end{equation}

\begin{equation}
\int_{0}^{T/2} i_3 \,dt = \int_{0}^{T/2} i_6 \,dt,
\label{eq_charge3}
\end{equation}

\noindent
where $T=1/f$ is the switching period ($f$ is the switching frequency), $r$ is the switch resistance, and $C_i$ is the flying capacitance.
Once the unknown current values are determined, the average output current for each stage can be determines as
\begin{equation}
I_{OUT1}=-\frac{1}{T}\left(\int_{0}^{T/2} i_1 \,dt + \int_{0}^{T/2} i_4 \,dt\right);
\label{IOUT1}
\end{equation}

\begin{equation}
I_{OUT2}=-\frac{1}{T}\left(\int_{0}^{T/2} i_2 \,dt + \int_{0}^{T/2} i_5 \,dt\right);
\label{IOUT2}
\end{equation}

\begin{equation}
I_{OUT3}=-\frac{1}{T}\left(\int_{0}^{T/2} i_3 \,dt + \int_{0}^{T/2} i_6 \,dt\right).
\label{IOUT3}
\end{equation}

The Y-parameters are determined by using (\ref{eq_Yij}). 
The R-parameters can then be obtained by inversion of Y-parameters matrix (\ref{eq_RandY_parameters}). 
After performing symbolic calculations using MATLAB, the simplified expression of FSL R-parameters matrix can be written as
\begin{equation}
R_{FSL} = 
\begin{bmatrix}
R_{par}+2r & R_{par}/2 & R_{par}/2\\
R_{par}/2 & 3R_{par}/2+2r & R_{par}\\
R_{par}/2 & R_{par} & 2R_{par}+2r
\end{bmatrix}
.
\label{eq_RFSL}
\end{equation}

From (\ref{eq_RFSL}), the non-diagonal terms are the functions of corresponding parasitic resistance values. 
This means that the coupling effect is only due to the parasitic resistance and the switch resistance values do not have an impact on the coupling. 
%\selcuk{why is this the case? do we have a justification?} 
Such observation can be explained with the fact that SC converter stages share only the parasitic resistance between each other. In other words, the charge that is provided to each load is flowing through certain parasitic resistances that are common to some or all loads depending on the configuration. In contrast, the charge that is flowing to the particular load depends only on the switch resistance of that load (\textit{i.e.}, it does not depend on the switch resistance of other loads because the charge is not flowing through them).
Furthermore, it can be observed that $R_{12}= R_{21},R_{13}= R_{31},R_{23}= R_{32}$. 
This means that the pair of outputs/loads affect each other in the equal amount. 
Therefore, the covert communication may exist in two directions.
Additionally, (\ref{eq_RFSL}) shows that the cross-regulation/coupling effects between the second and third loads are two times larger than in the first-second and the first-third pairs. 
%\selcuk{why is this the case? what is the importance of this?}
%\selcuk{also, please refrain from using one-sentence paragraphs.} 
This behavior is caused by the location of the SC converter stages with respect to the input voltage source. 
Since the second and third loads are located the farthest from the input voltage source, the amount of coupling is the largest between them. 
The larger amount of coupling corresponds to a larger bandwidth of covert communication. 
Therefore, it is important to find the pair of loading conditions that lead to the largest coupling to maximize the bandwidth of covert communication channel.

\subsubsection{SSL}
\begin{figure}[t]
	\centering
	\includegraphics[width=0.4\textwidth]{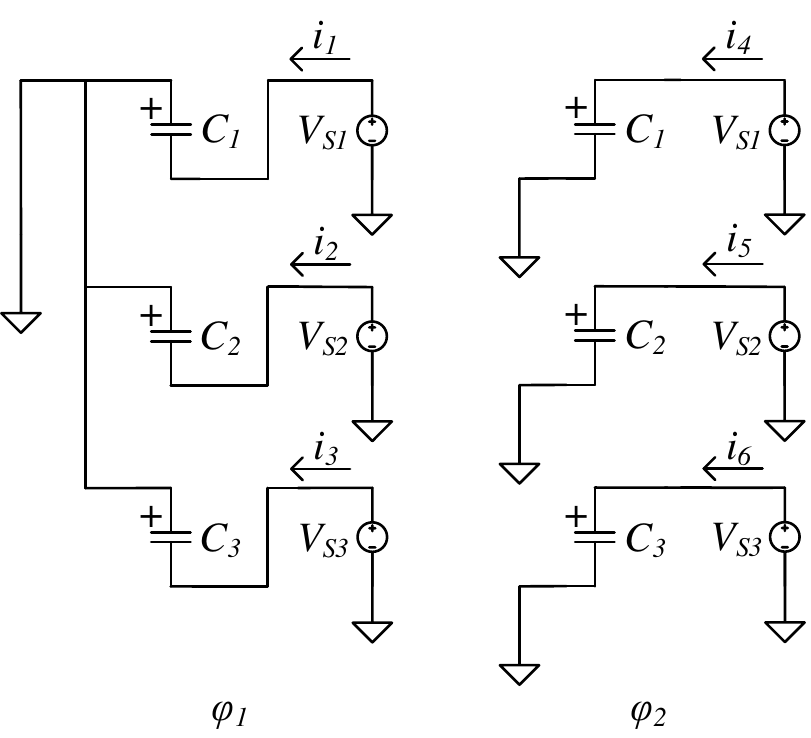}
	\caption{SSL Y-parameter calculation for a three-stage 2:1 SC converter supplying three different loads.}
	\label{fig:SSL_3-port_network}
\end{figure}

The circuit diagram used to determine SSL Y-parameters is shown in Fig. \ref{fig:SSL_3-port_network}. 

Since there are no resistances in the SSL, the average current through a capacitor that flows to the positive terminal can be written as
\begin{equation}
I_C=\frac{C(V_C\{t_2\}-V_C\{t_1\})}{t_2-t_1},
\label{eq_capacitor_current}
\end{equation}
\noindent
where $V_C\{t\}$ is the capacitor voltage at time $t$.
From Fig. \ref{fig:SSL_3-port_network}, the current values can be expressed as
\begin{equation}
i_1=i_4=\frac{4CV_{S1}}{T};
\label{eq_i1andi4}
\end{equation}

\begin{equation}
i_2=i_5=\frac{4CV_{S2}}{T};
\label{eq_i2andi5}
\end{equation}

\begin{equation}
i_3=i_6=\frac{4CV_{S3}}{T}.
\label{eq_i3andi6}
\end{equation}

The remaining parameters such as the average output currents, and Y- and R-parameters can be determined similar to the FSL case (\ref{eq_RandY_parameters}), (\ref{eq_Yij}), (\ref{IOUT1})-(\ref{IOUT3}). 
The simplified expression of SSL R-parameters matrix can be written as
\begin{equation}
R_{SSL} = 
\begin{bmatrix}
\frac{1}{4Cf} & 0 & 0\\
0 & \frac{1}{4Cf} & 0\\
0 & 0 & \frac{1}{4Cf}
\label{eq_RSSL}
\end{bmatrix}.
\end{equation}

As shown in (\ref{eq_RSSL}), all non-diagonal terms are zero. 
This means that during the SSL operation mode (low switching frequency), there is no coupling effect between loads.

\subsection{Simulation-based modeling}

\begin{table}[t]
\caption{Simulation parameters.}
\begin{center}
\begin{tabular}{|c|c|}
        \hline
		\textbf{Parameter} & \textbf{Value}\\
		\hline
		Input voltage $V_{IN}$ & 1 $V$\\
		\hline
		Flying capacitance $C_1,C_2,C_3$ & 1 $\mu F$\\
		\hline
		Filter/load capacitance $C_{O1},C_{O2},C_{O3}$ & 10 $\mu F$\\
		\hline
		Switch resistance $r$ & 0.1 $\Omega$\\
		\hline
		Parasitic resistance $R_{par}$ & 0.01 $\Omega$\\
		\hline
		Switching frequency $f$ & 10 $MHz$\\
		\hline
\end{tabular}
\label{table:1}
\end{center}
\end{table}

The parameter values used in the simulation-based modeling of the three-stage SC converter are listed in Table \ref{table:1}. Since there are three outputs/loads, the circuit can be viewed as a 3-port network. 
The dimensions of R-parameters matrix is therefore three by three (\textit{i.e.}, in total, there are nine elements). 
To determine each individual element of the R-parameters matrix, four measurements are required. 
The first measurement is used to determine $V_{TRi}$ (\ref{eq_target_VOUT}). 
The remaining three measurements are used to determine the three columns of R-parameters matrix. 
Please note that non-overlapping switching is used to eliminate the unwanted capacitor discharge, which is a common practice in the design of SC converters~\cite{saiz2006low}. 
%\selcuk{can you put a reference here for the non overlapped switching case?} \yerzhan{\{Reference is added.\}}

By following the measurement methodology in Section \ref{sim_modeling}, the R-parameters are determined as
\begin{equation}
R = 
\begin{bmatrix}
216.9 & 5.107 & 5.112\\
5.102 & 222.0 & 10.21\\
5.113 & 10.21 & 227.1
\end{bmatrix}
m\Omega
\label{eq_R_simulated}
\end{equation}
A switching frequency of 10 $MHz$ is used to enter FSL operation mode. 
For the case when $r=0.1 \: \Omega$ and $R_{par}=0.01 \:\Omega$ parameters, (\ref{eq_RFSL}) can be expressed as
\begin{equation}
R_{FSL} = 
\begin{bmatrix}
210 & 5 & 5\\
5 & 215 & 10\\
5 & 10 & 220
\end{bmatrix}
m\Omega
\label{eq_RFSL_values}
\end{equation}

The simulated results are in good agreement with analytical model (\ref{eq_RFSL_values}). 
One may argue that the diagonal terms have larger difference between analytical model and simulated results than non-diagonal terms. 
%\{This is due to the impact of SSL component (\ref{eq_RSSL}), though it is small due to the high frequency.\}
%\selcuk{can we rewrite this last sentence?}
This difference is caused by the impact of SSL component (\ref{eq_RSSL}) that increases the values of R-parameters. The SSL component is non-zero because the finite switching frequency of 10 $MHz$ is used.

\section{Simulation of Covert Communication}\label{simulation_covert_communication}

A source entity powered by one of the SC converter stages can encode information (bits) by varying its own workload.
For example, when the load current in the source entity increases, the corresponding SC converter stage needs to increase the output current to mitigate any potential  drop at the output voltage level.
These changes in the source eventually couples to the sink entity through the PDN and SC converter.
%By varying the load of one stage, it is possible to encode the information (bits) and transmit it to other loads. 
According to (\ref{eq_generalized_multi_port}), to achieve the maximum coupling effect between loads, the non-diagonal elements of R-parameters matrix need to be maximized and load/output current needs to be controlled in such a way that the output current at the source is maximum, whereas the output current at sinks is minimum.
The latter is done to minimize the noise coming from the sinks.

To establish a covert communication channel, the same circuit used in the case study is chosen (Fig. \ref{fig:3-port_network}). 
One way to control the output current is to vary the load resistance. 
By setting the load resistance to 100 $\Omega$, the output current becomes very small in the order of $\mu A$. 
Similarly, at the load resistance of 1 $\Omega$, the output current is in the order of $mA$. These values are specifically chosen for the circuit parameters (Table \ref{table:1}). 
Let us assume that bit `1' is encoded when the load resistance is 100 $\Omega$ while bit `0'  is encoded when the load resistance is 1 $\Omega$. 
Alternatively, a `1' is sent by reducing the workload and therefore current consumption and `0' is sent by increasing the workload in the source entity.
At the same time, the sink is assumed to have a load resistance of 100 $\Omega$. 
A covert communication throughput of 40 $kbits/s$ is obtained in the simulations with a potentially higher bandwidth.
%The communication throughput is chosen to be 40 $kbits/s$. 
The bandwidth depends on the response time of SC converters. 
If the communication throughput is selected to be higher than the channel bandwidth, SC converters will not be able to respond to the change in load, and in turn no meaningful communication would be established between the source and sink.

\begin{figure}[t]
	\centering
	\includegraphics[width=0.5\textwidth]{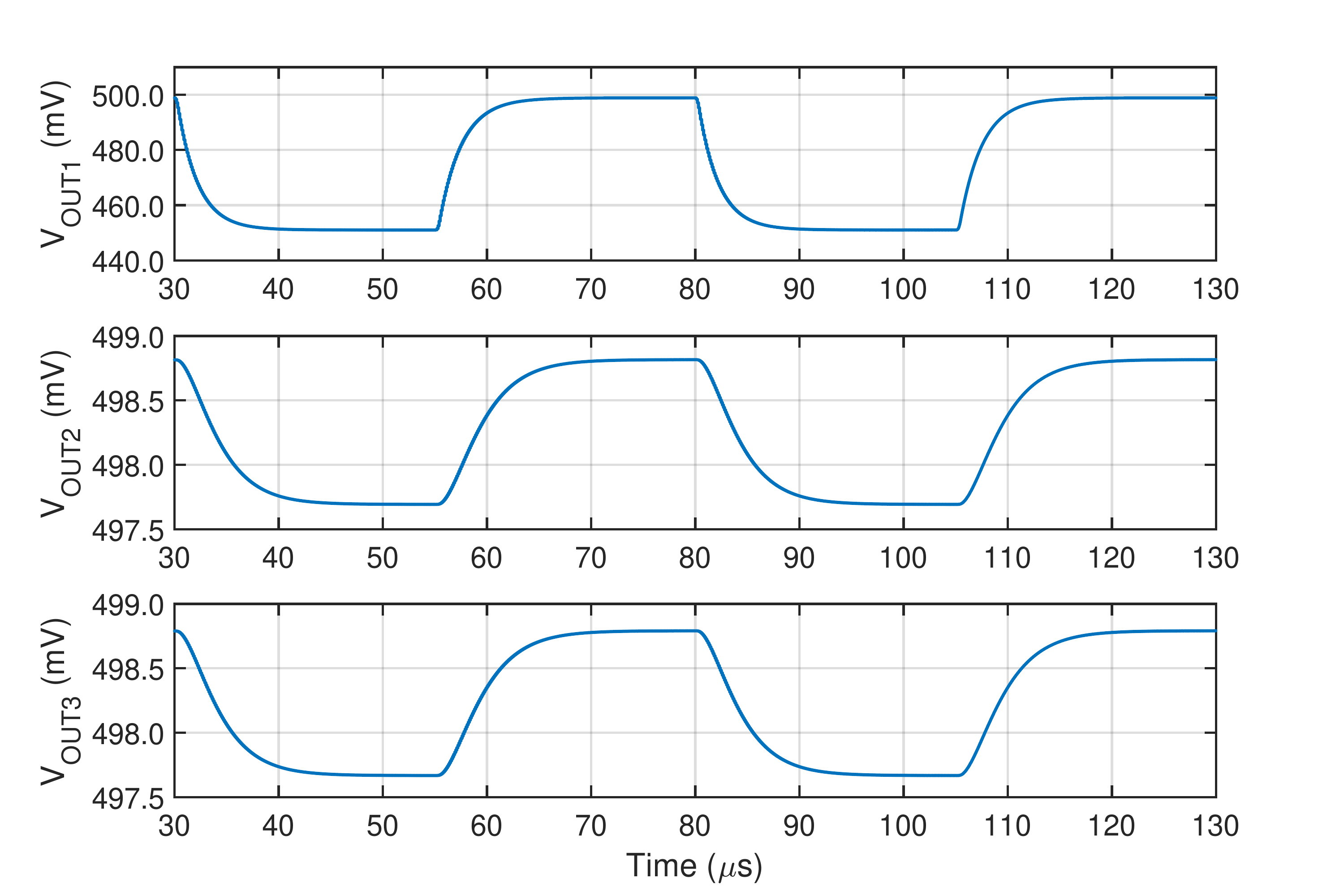}
	\caption{Output voltage waveforms at the switching frequency of 10 $MHz$. The information is sent from the first SC converter stage by varying $R_1$.}
	\label{fig:covert_comm_10MHz_traces_Vout}
\end{figure}

\begin{figure}[t]
	\centering
	\includegraphics[width=0.5\textwidth]{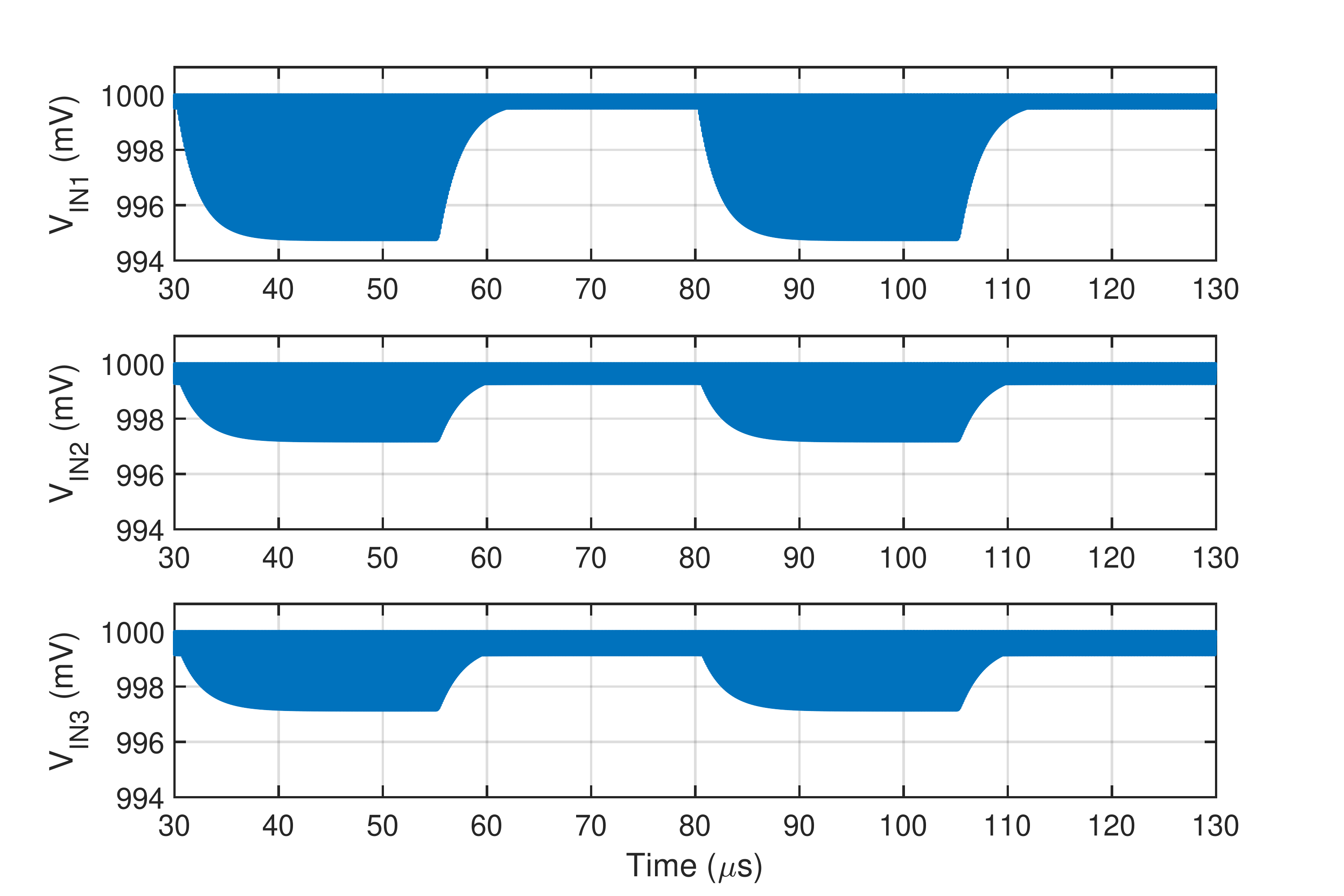}
	\caption{Input voltage waveforms at the switching frequency of 10 $MHz$. The information is sent from the first SC converter stage by varying $R_1$.}
	\label{fig:covert_comm_10MHz_traces_Vin}
\end{figure}

The output and input voltage waveforms for each load at a switching frequency of 10 $MHz$ (FSL mode) are shown in Figs. \ref{fig:covert_comm_10MHz_traces_Vout} and \ref{fig:covert_comm_10MHz_traces_Vin}. 
In this case, the information is sent from the  source by changing the value of load $R_1$. The signal propagation from the source to the sinks can be observed from these figures. For example, in Fig. \ref{fig:covert_comm_10MHz_traces_Vout}, the amplitude variation of signal at the source side (\textit{i.e.}, at the output of first SC converter stage) is approximately 47.9 $mV$, whereas the amplitude variation of the received signal is 1.12 $mV$ at the second and third SC converter stages.

\begin{figure}[t]
	\centering
	\includegraphics[width=0.5\textwidth]{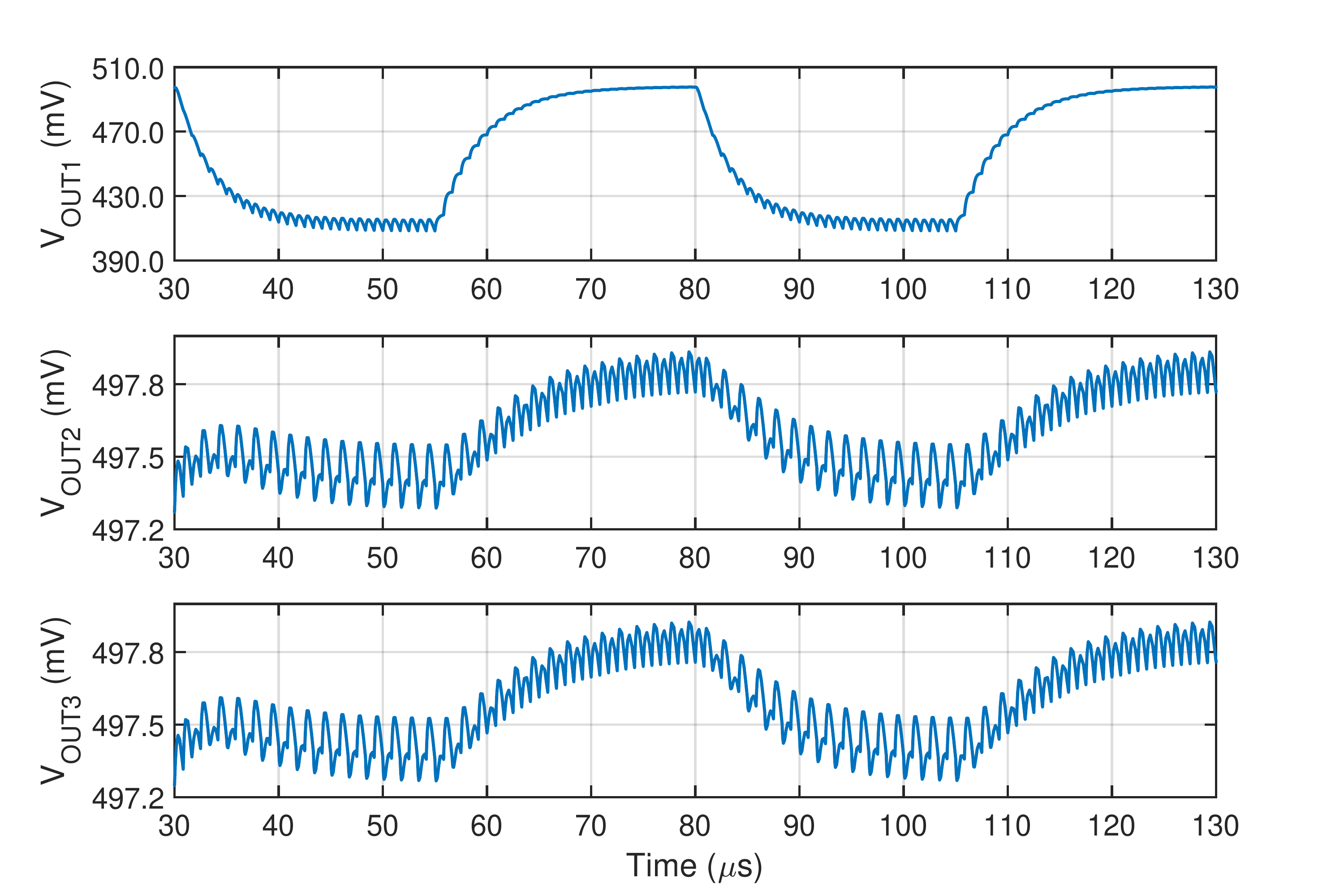}
	\caption{Output voltage waveforms at the switching frequency of 600 $kHz$. The information is sent from the first SC converter stage by varying $R_1$.}
	\label{fig:covert_comm_600kHz_traces_Vout}
\end{figure}

\begin{figure}[t]
	\centering
	\includegraphics[width=0.5\textwidth]{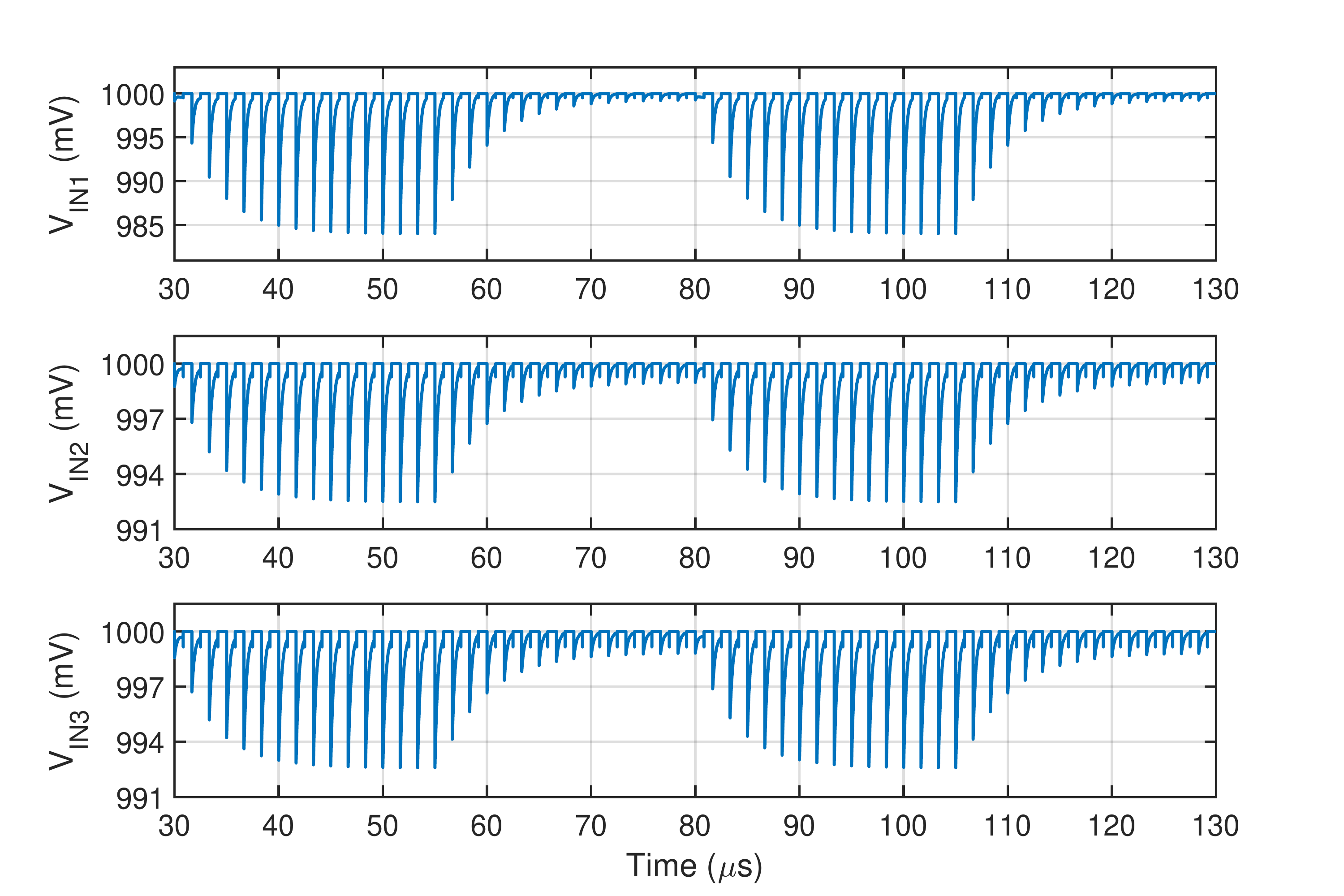}
	\caption{Input voltage waveforms at the switching frequency of 600 $kHz$. The information is sent from the first SC converter stage by varying $R_1$.}
	\label{fig:covert_comm_600kHz_traces_Vin}
\end{figure}

At a significantly lower switching frequency of 600 $kHz$, the variations in load become more noisy because of the increased voltage ripples (Figs. \ref{fig:covert_comm_600kHz_traces_Vout} and \ref{fig:covert_comm_600kHz_traces_Vin}). 
Additionally, the response time of the SC converter reaches the limit of communication throughput (\textit{i.e.}, if the switching frequency of SC converter is further reduced, the SC converter will not be able to respond to the change in load). 
%\selcuk{so what?, NEED TO DISCUSS!!} \yerzhan{Therefore, as the switching frequency of SC converter decreases, both communication strength as well as bandwidth are reduced. }

\begin{figure}[t]
	\centering
	\includegraphics[width=0.5\textwidth]{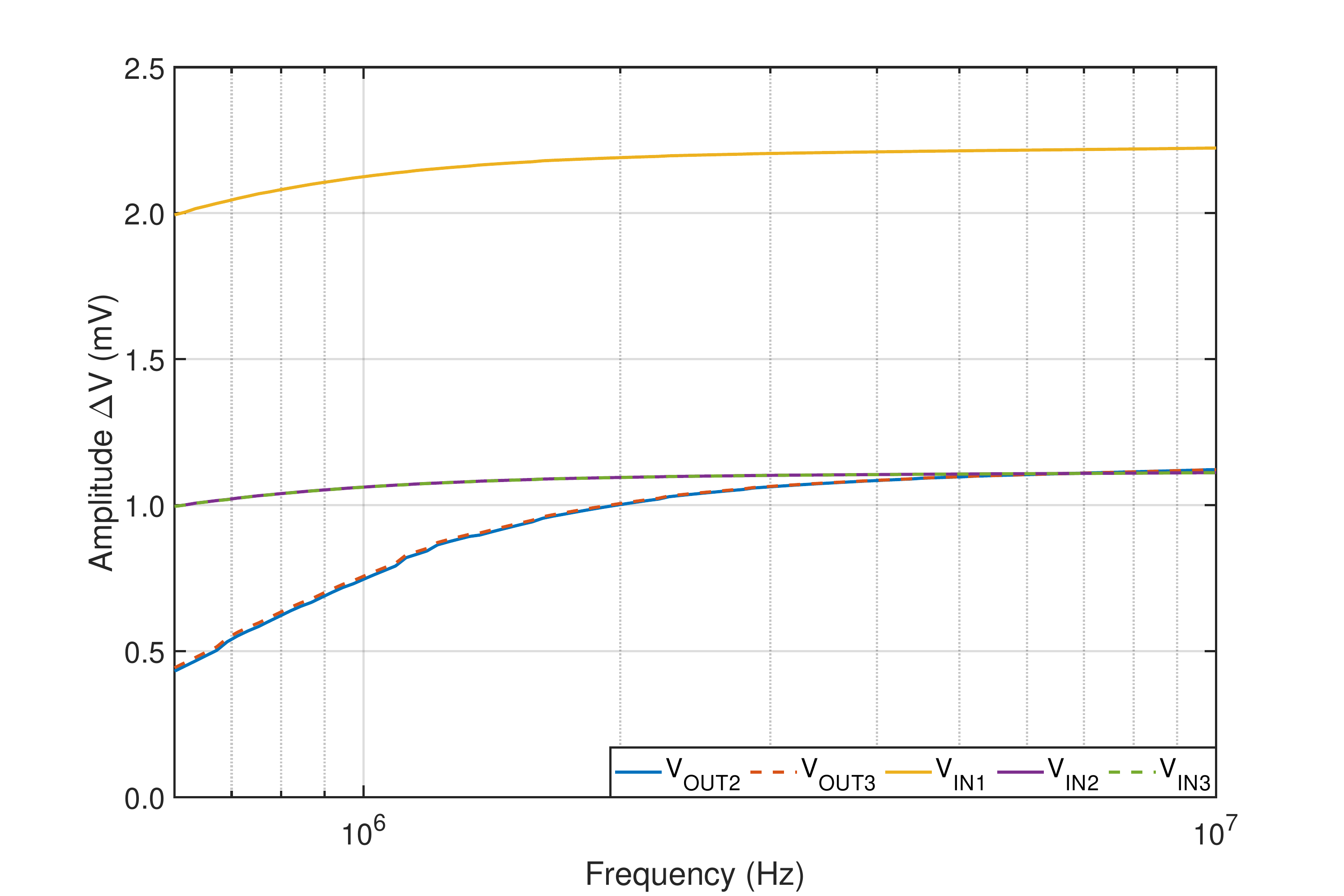}
	\caption{The amplitude variations of voltage at different nodes. The information is sent from the first SC converter stage by varying $R_1$.}
	\label{fig:covert_comm_source_1_traces}
\end{figure}

\begin{figure}[t]
	\centering
	\includegraphics[width=0.5\textwidth]{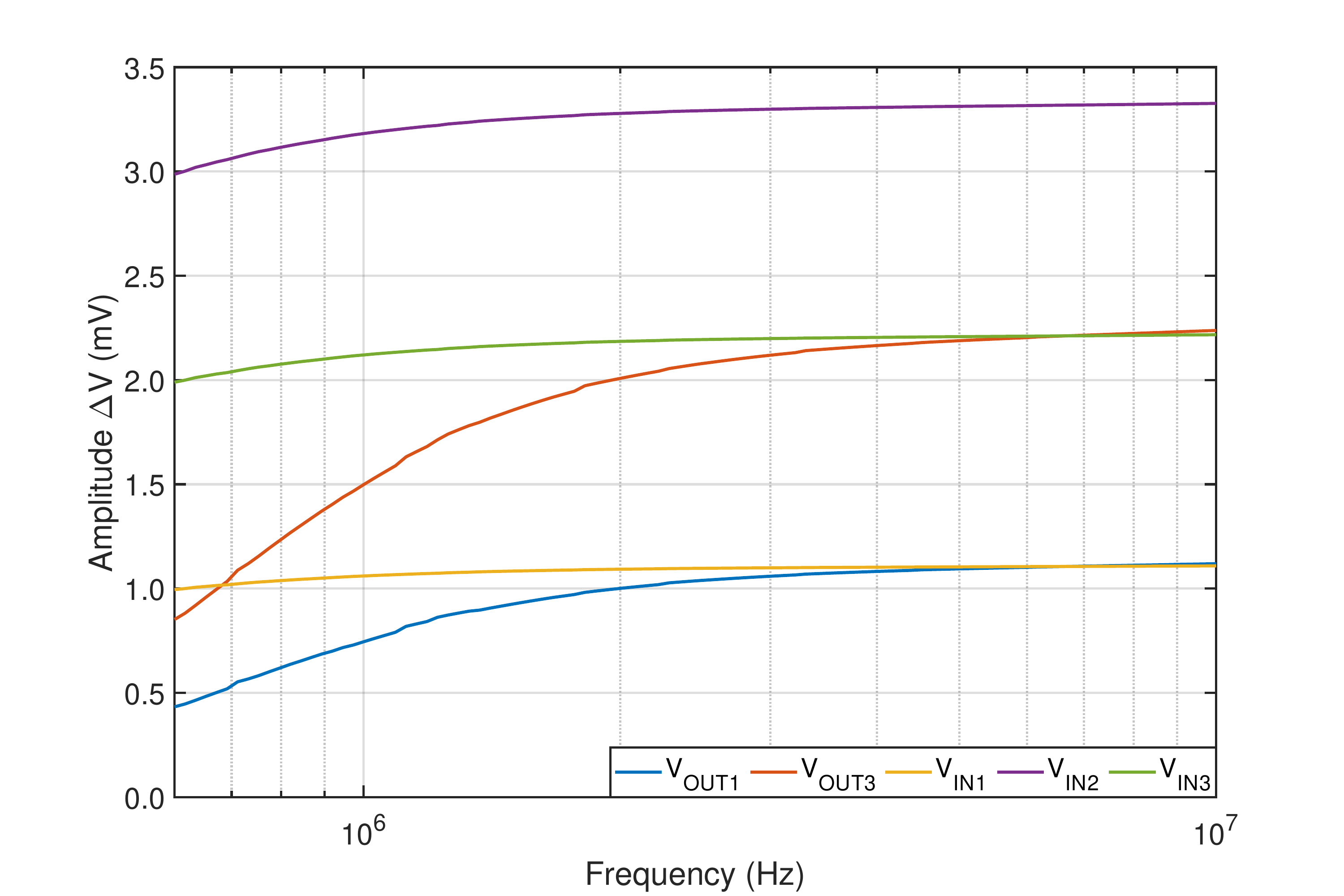}
	\caption{Amplitude variations of voltage at different nodes. The information is sent from the second SC converter stage by varying $R_2$.}
	\label{fig:covert_comm_source_2_traces}
\end{figure}

\begin{figure}[t]
	\centering
	\includegraphics[width=0.5\textwidth]{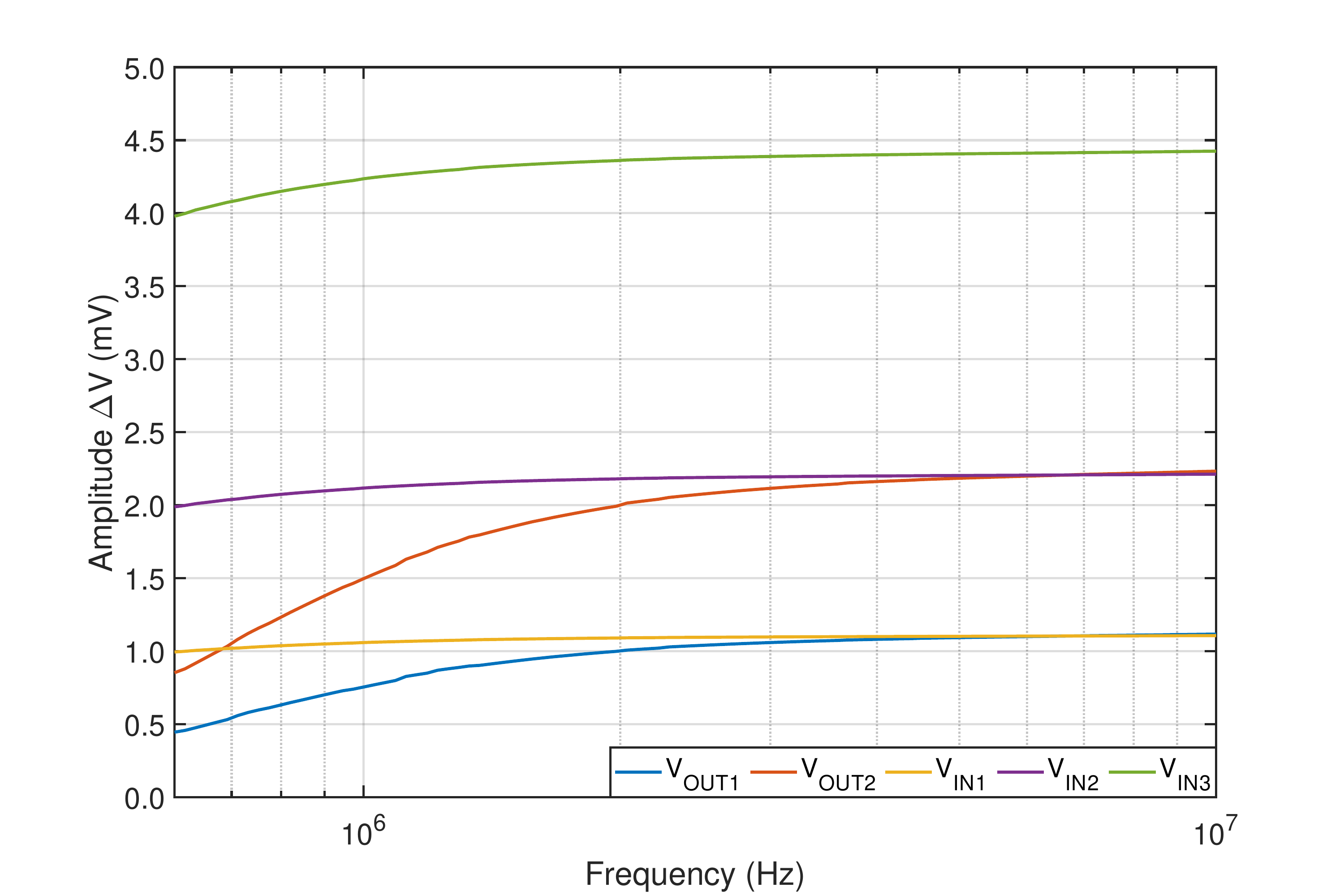}
	\caption{Amplitude variations of voltage at different nodes. The information is sent from the third SC converter stage by varying $R_3$.}
	\label{fig:covert_comm_source_3_traces}
\end{figure}

To quantify the coupling effects and communication bandwidth, the amplitude variation of the voltage ($\Delta V$) is measured. 
%\selcuk{instead of communication strength, shall we use communication bandwidth?} \yerzhan{\{The coupling shows the impact of average output voltage, which is the communication strength. The bandwidth depends only on the response time of SC converter.\}}
%Particularly, the average voltage is measured separately for both bit `0' and bit `1'.
%\selcuk{where do you measure the average voltage?}
After that, the change in two values is recorded. 
%\selcuk{be more explicit.for example: the amplitude of the voltage at node YYY is measured, respectively, when a bit 1 and bit 0 is transmitted}\}
Particularly, the average voltage at the output and input nodes of SC converter stages is measured over the time period when the bit `0' and bit `1' are transmitted. Please note that this time period corresponds to the steady-state operation region of SC converter (\textit{i.e.}, the response/transient time is not considered).
The amplitude variation of voltages at various nodes when the information is sent from load $R_1$, $R_2$, $R_3$ is shown in Figs. \ref{fig:covert_comm_source_1_traces}-\ref{fig:covert_comm_source_3_traces}, respectively. It should be noted that the amplitude variation at the source side is not included in these figures because of significantly larger amplitude value as is shown in Figs. \ref{fig:covert_comm_10MHz_traces_Vout}-\ref{fig:covert_comm_600kHz_traces_Vin}.
The relationship between the amplitude variations and the switching frequency of SC converter can be observed from Figs. \ref{fig:covert_comm_source_1_traces}-\ref{fig:covert_comm_source_3_traces}. 
%\{The amplitudes saturate with high switching frequency.\} 
%\selcuk{rewrite} 
The amplitude $\Delta V$ for each node increases with switching frequency of SC converter and then saturates at FSL region.
This is because SC converter stages enter the FSL operation mode, where the coupling effects are the largest (\ref{eq_RFSL}). 
As the switching frequency decreases, the coupling effects are reduced. 
It should be noted that the rate of decrease is larger in the output voltage nodes compared to the input voltage nodes. This behavior is caused by the amount of voltage drop across a particular SC converter stage. At a high switching frequency (\textit{i.e.}, FSL mode), the voltage drop across an SC converter is the lowest because the equivalent resistance is minimum \cite{seeman2009design,mustafa2018scc}. This means that the variations in the input and output voltages of an SC converter stage are approximately equal. As the switching frequency decreases, the equivalent resistance increases, making the voltage drop across the SC converter stage larger. As a result, the variations in the input voltage cause smaller variation in the output voltage due the larger voltage drop. Such behavior can be observed in Figs. \ref{fig:covert_comm_source_1_traces}-\ref{fig:covert_comm_source_3_traces}.
%
%\selcuk{do we have an answer for this question?} \yerzhan{\{I have some ideas. Let's discuss them during the meeting. Briefly, it might be related to the voltage drop across the SC converter stage. For example, if the input voltage drops by half and it is FSL, the same voltage drop will be at output because equivalent resistance is much smaller than the load resistance. If it is SSL, then there will be higher voltage drop across the SC converter stage and therefore lower change at output.\}}
%
Therefore, the level of coupling effects can be modified by simply varying the switching frequency of SC converters.

%\{From Fig. \ref{fig:covert_comm_source_1_traces}, the load $R_1$ affects $V_{OUT2}$ and $V_{OUT3}$ in the same amount. The similar behavior holds for the input voltage nodes $V_{IN2}$ and $V_{IN3}$.\} 
From Fig. \ref{fig:covert_comm_source_1_traces}, the changes in workload of the source entity located in the first SC converter stage (\textit{i.e.}, $R_1$) induces the same changes in voltage of sink entities located in the second and third SC converter stages. Particularly, $\Delta V_{OUT2}=\Delta V_{OUT3}$ and $\Delta V_{IN2}=\Delta V_{IN3}$. This can be explained with the fact that $R_{21}=R_{31}$ in (\ref{eq_RFSL}).

%\{From Fig. \ref{fig:covert_comm_source_2_traces}, the load $R_2$ affects $V_{OUT3}$ approximately two times larger than $V_{OUT1}$. \selcuk{instead of writing as is $R_2$ affects Vout, we should write that the changes in the source entity (in this case changes in the workload of second SC converter stage) induces some differences in the sink entities.}
%The similar behavior holds for the input voltage nodes $V_{IN3}$ and $V_{IN1}$.\}

From Fig. \ref{fig:covert_comm_source_2_traces}, the changes in workload of the source entity located in the second SC converter stage (\textit{i.e.}, $R_2$) induces approximately two times larger changes in voltage of the sink located in the third SC converter stage compared to voltage changes of the sink located in the first SC converter stage. Particularly, $\Delta V_{OUT3}=2\Delta V_{OUT1}$ and $\Delta V_{IN3}=2\Delta V_{IN1}$. This is due to the fact that $R_{32}=2R_{12}$ in (\ref{eq_RFSL}). With a similar reasoning (\textit{i.e.}, $R_{23}=2R_{13}$), the changes in workload of the source entity located in the third SC converter stage (\textit{i.e.}, $R_3$) results in the following relation: $\Delta V_{OUT2}=2\Delta V_{OUT1}$ and $\Delta V_{IN2}=2\Delta V_{IN1}$ (Fig. \ref{fig:covert_comm_source_3_traces}). 

Therefore, the developed analytical model (\ref{eq_RFSL}) can successfully predict the relative impact of one entity on another entity. 
In this way, an adversary may choose the locations of the source and sink such that the coupling effects are maximized. 
In this case study, the highest coupling effects exist between the second and third SC converter stages. 

\begin{figure}[t]
	\centering
	\includegraphics[width=0.5\textwidth]{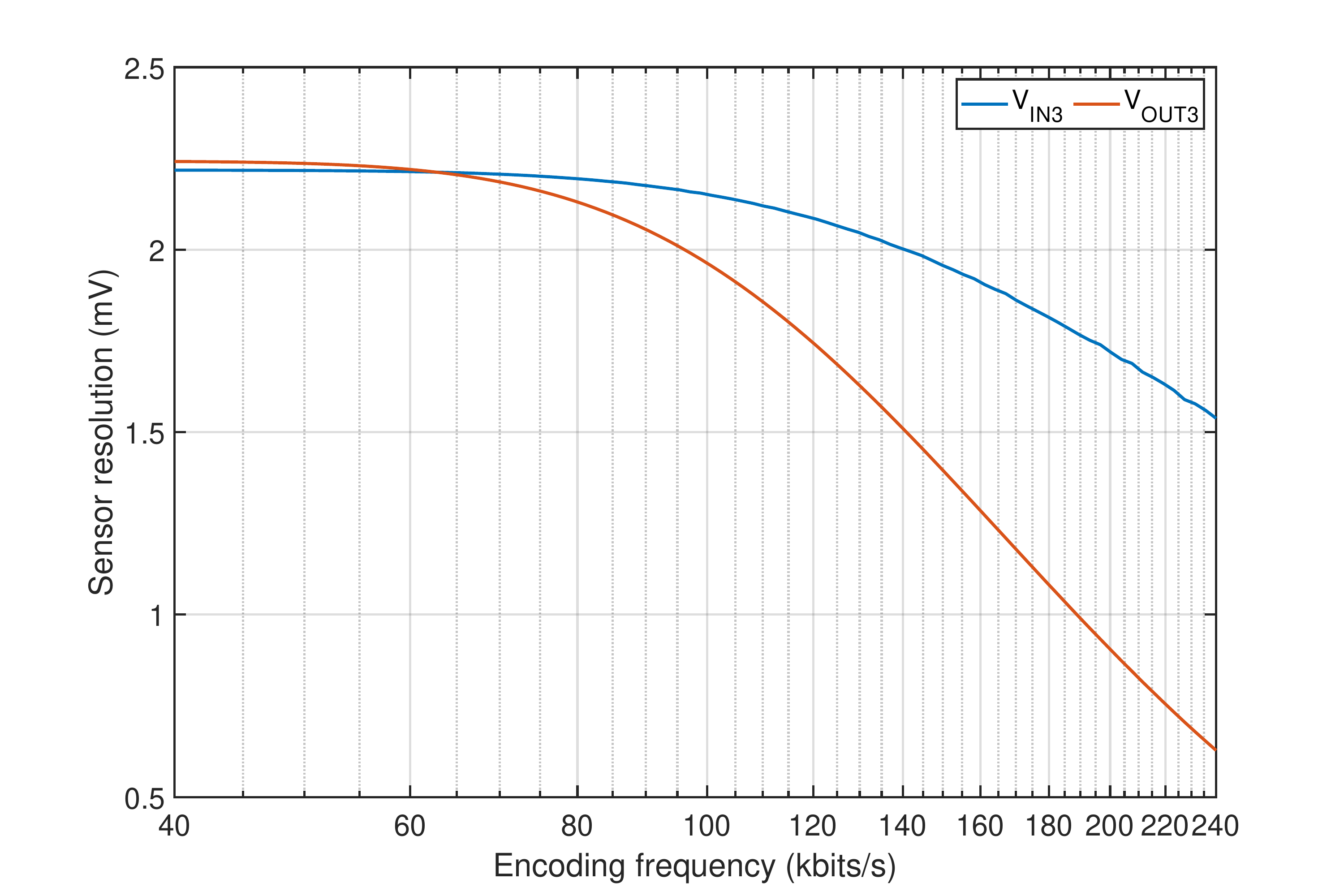}
	\caption{Amplitude variations of voltage at the input and output nodes of the third SC converter stage. The information is sent from the second SC converter stage.}
	\label{fig:covert_comm_10MHz_2to3_bandwidth}
\end{figure}

Once the pair of SC converter stages that exhibit the largest coupling effects is identified, the bandwidth of the covert communication channel is analyzed. 
The sensor resolution that is needed to detect the variation of input and output voltages of the third SC converter stage as a function of the encoding frequency is shown in Fig. \ref{fig:covert_comm_10MHz_2to3_bandwidth}. 
In this case, the information is transmitted from the second SC converter stage. 
The switching frequency of SC converter is set to 10 $MHz$ (\textit{i.e.}, FSL mode) to achieve the largest coupling effects. 
As can be observed, at the input node, the bandwidth is larger as compared to the output node. 
This is due the effect of SC converter response time. 
In Figs. \ref{fig:covert_comm_10MHz_traces_Vout} and \ref{fig:covert_comm_10MHz_traces_Vin}, one may notice that the response time at the input node is smaller than at the output node. 
The value of bandwidth depends on the resolution of sensor that detects changes in voltage. 
For example, if the resolution is equal to 2 $mV$, the bandwidth is approximately equal to 95 $kbits/s$ and 140 $kbits/s$ for the output and input nodes, respectively (Fig. \ref{fig:covert_comm_10MHz_2to3_bandwidth}). 

\begin{figure}[t]
	\centering
	\includegraphics[width=0.4\textwidth]{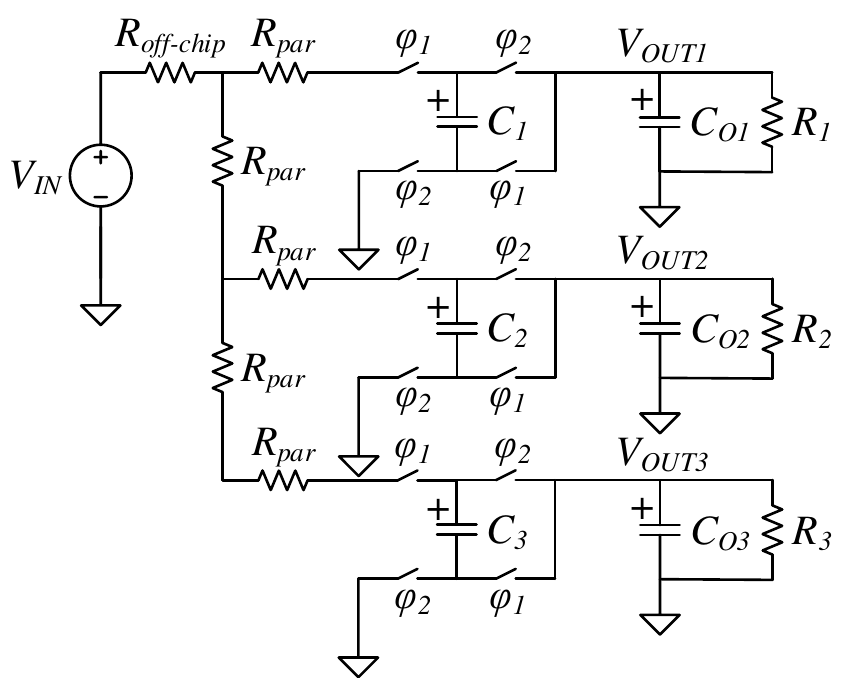}
	\caption{Three-stage 2:1 SC converter supplying three different loads with off-chip resistance $R_{off-chip}$.}
	\label{fig:3-port_network_with_off-chip_R}
\end{figure}

In order to further increase the coupling effects, an adversary may connect an off-chip resistance to the supply voltage (Fig. \ref{fig:3-port_network_with_off-chip_R}) to increase the effective impedance between the on-chip SC converter and the off-chip power supply.
Intuitively, when the off-chip resistance increases, the changes induced by one of the SC converter stages to the input node of the other SC converter stages cannot be mitigated as strongly by the off-chip power supply as time constant between is increased.
The implication of increasing the off-chip resistance on the coupling among SC converter stages is investigated both with the analytical model and simulations.

\begin{figure}[t]
	\centering
	\includegraphics[width=0.5\textwidth]{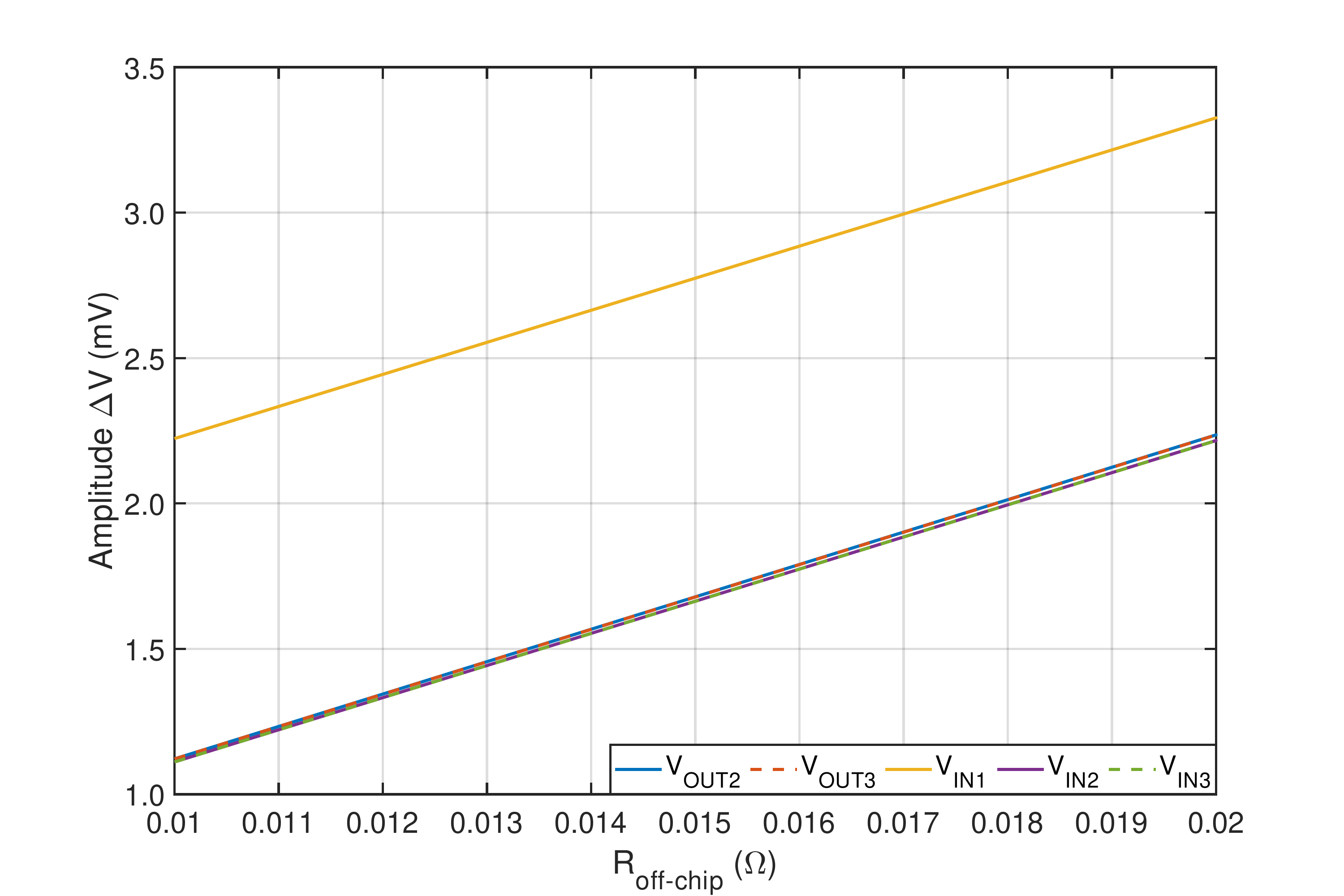}
	\caption{Amplitude variations of voltage as the function of $R_{off-chip}$ at the switching frequency of 10 $MHz$. The information is sent from the first SC converter stage by varying $R_1$.}
	\label{fig:covert_comm_source_1_traces_Roffchip}
\end{figure}

By following the same approach offered in Section \ref{analytical_modeling}, the FSL R-parameters matrix is determined as 
\begin{equation}
\begin{bmatrix}
R_{FSL11}\\
R_{FSL21}\\
R_{FSL31}
\end{bmatrix}
= 
\begin{bmatrix}
\frac{R_{off-chip}}{2}+\frac{R_{par}}{2}+2r\\
\frac{R_{off-chip}}{2}\\
\frac{R_{off-chip}}{2}
\end{bmatrix}
;
\label{eq_RFSL_Roffchip1}
\end{equation}
\begin{equation}
\begin{bmatrix}
R_{FSL12}\\
R_{FSL22}\\
R_{FSL32}
\end{bmatrix}
= 
\begin{bmatrix}
\frac{R_{off-chip}}{2}\\
\frac{R_{off-chip}}{2} + R_{par} + 2r\\
\frac{R_{off-chip}}{2} + \frac{R_{par}}{2}
\end{bmatrix}
;
\label{eq_RFSL_Roffchip2}
\end{equation}
\begin{equation}
\begin{bmatrix}
R_{FSL13}\\
R_{FSL23}\\
R_{FSL33}
\end{bmatrix}
= 
\begin{bmatrix}
\frac{R_{off-chip}}{2}\\
\frac{R_{off-chip}}{2} + \frac{R_{par}}{2}\\
\frac{R_{off-chip}}{2} + \frac{3R_{par}}{2}+2r
\end{bmatrix}
.
\label{eq_RFSL_Roffchip3}
\end{equation}
There is a term $R_{off-chip}/2$ in each matrix element (including non-diagonal elements) in (\ref{eq_RFSL_Roffchip1})-(\ref{eq_RFSL_Roffchip3}). 
Due to the existence of this term, the off-chip resistance has an equal effect on the coupling between different entities.
The simulation results of amplitude variations as a function of the off-chip resistance at the switching frequency of 10 $MHz$ are shown in Fig. \ref{fig:covert_comm_source_1_traces_Roffchip}. 
In this case, the encoding in the source is performed by changing the load $R_1$. 
%\selcuk{be more explicit. what amplitude?} 
There is a linear relationship between the off-chip resistance and the amplitude variation of voltages at various nodes when the bit `0' and bit `1' are transmitted.

\section{Conclusion}\label{conclusion}
In this work, a novel covert communication channel between SC converter stages which share the same PDN is introduced. 
This vulnerability could be used to establish an inter- or intra-core covert communication channel. 
By using the proposed multi-port network modeling technique, the bandwidth of covert communication channel between different entities is analyzed. 
In particular, the impact of circuit components such as switch resistance, capacitance, switching frequency, parasitic resistance, off-chip resistance, and location of source and sink is investigated with the proposed analytical model. 
This model can be used both by an attacker to increase the coupling or a circuit designer to minimize the coupling. 
As a case study, a three-stage 2:1 SC converter is used to demonstrate the effectiveness of the proposed modeling technique. 
The analytical model is verified with extensive simulations.

%As a future work, it is possible to generalize the modeling technique for other types of VRs such as buck converter and low-dropout regulator. 

%\section*{Acknowledgment}

\bibliographystyle{./bibliography/IEEEtran}
\bibliography{./bibliography/IEEEabrv,./bibliography/IEEEexample}

\end{document}